\documentclass[prl,aps,onecolumn,amsmath,amssymb,letterpaper,12pt]{revtex4}
\usepackage{hyperref}
\pdfoutput=1
\usepackage{graphic x}
\usepackage{footnote}
\usepackage{amsmath}
\usepackage{bm}

\def\beq{\begin{equation}}
\def\eeq{\end{equation}}
\def\bea{\begin{eqnarray}}
\def\eea{\end{eqnarray}}

\def\kt{k_{\rm B} T}

\begin{document}

\title{Self-assembly of amphiphilic peanut-shaped nanoparticles}
\author{Stephen Whitelam\footnote{swhitelam@lbl.gov}$^1$ and Stefan A.F. Bon$^2$} 
\affiliation{$^{1}$Molecular Foundry, Lawrence Berkeley National Laboratory, 1 Cyclotron Road, Berkeley, CA 94720, USA\\
$^{2}$Department of Chemistry, University of Warwick, Coventry, CV4 7AL, United Kingdom}

\begin{abstract}
We use computer simulation to investigate the self-assembly of Janus-like amphiphilic peanut-shaped nanoparticles, finding phases of clusters, bilayers and micelles in accord with ideas of packing familiar from the study of molecular surfactants. However, packing arguments do not explain the hierarchical self-assembly dynamics that we observe, nor the coexistence of bilayers and faceted polyhedra. This coexistence suggests that experimental realizations of our model can achieve multipotent assembly of either of two competing ordered structures.
\end{abstract}
\maketitle
Components are said to `self-assemble' when they organize to form stable patterns or aggregates without external direction. Self-assembly is driven by interactions as different as weak covalent bonds and capillary forces, involves components ranging in size from ${\textnormal \AA}$ngstr\"oms to centimeters, and occurs both in inorganic settings and in living organisms~\cite{tien1998cms,mao2002dsa,breen1999das,whitesides2002bms,whitesides,rabani2003dms,boal2000san,dobson2003pfa,virus_ref}. Mimicry of the self-assembly seen in the natural world promises the development of new, functional materials patterned on the nanometer scale~\cite{huie2003gms,zhang2003fnb}. In pursuit of this goal, we take inspiration from the self-assembly of molecular surfactants to investigate using computer simulation the behavior of a simple model of their colloidal counterparts. Molecular surfactants, comprising chemically linked hydrophobic and hydrophilic groups, are of central importance in biology and industry, able to form a plethora of phases in water or mixtures of water and oily liquids. These phases include micelles~\cite{maibaum2004mfa}, bilayers and vesicles, as well as numerous bicontinous phases~\cite{schwarz2002bss} that serve as internal cellular packaging~\cite{landh1995eme} and are the bane of many a plumber.

Here we ask: What might self-assemble in aqueous solution from amphiphilic, peanut-shaped colloidal nanoparticles? Such particles can now be prepared in large quantities, starting from spherical, crosslinked polystyrene `seed' particles of radius $\sim50$ nm. Mixing these seeds with styrene initiates monomer-polymer phase separation and creates particles with a peanut-like shape characterized by two fused spherical lobes of controllable relative size. Additional treatment of the seed particle renders peanuts amphiphilic, with one lobe hydrophobic and the other hydrophilic~\cite{mock2006san,sheu1990psp,bon1}; the resulting body can be regarded as a generalized Janus particle~\cite{walther2008jp}. (We note that colloidal silica dumbbells can be synthesized by other routes~\cite{johnson2005scs}, and that micrometer-scale peanuts show potential as Pickering stabilizers of oil-in-water emulsions~\cite{kim2004snc}.) In an attempt to answer our posed question we have constructed a model of interacting peanuts whose minimal character is motivated by the insight into self-assembly afforded by similarly simple model systems~\cite{huisman2007ptb,Glotzer0,Glotzer1,patchy1,patchy2,mike,villar2009saa,ouldridge2009sad,sciortino2007self,bianchi2007fully,bianchi2006phase,de2006dynamics,chen2007pps,hong2008cac,mladek2008computer,elrad2008msc,iacovella2009ccs,miller2009hsa, sciortino2009phase}. Our model can be evolved with computational efficiency sufficient to allow observation of collective, thermally-driven dynamics on timescales of seconds. Its construction rests upon two assumptions: first, that peanuts in aqueous solution experience a thermodynamic driving force that causes hydrophobic lobes to attract each other; and second, that functionalization of peanuts' hydrophilic lobes renders them chemically passive. We assume solution conditions to be such that electrostatic interactions between peanuts mediate only short-ranged repulsions. 

Model geometry is shown in the Appendix (Fig.~\ref{figsupp2}). Each peanut consists of a spherical hydrophobic lobe of radius $R_0$ fused with a spherical hydrophilic lobe of radius $R_1$. The centers of the two lobes are separated by a distance $\delta$, which we quantify in dimensionless form via the parameter $\epsilon \equiv \delta/(R_0+R_1)$. We fixed $R_0$ throughout (we consider $R_0$ to be approximately 50 nm), and investigated the behavior of the model as we varied $\epsilon$ and $R_1/R_0$. We required $|R_1 -R_0|/(R_1+R_0) < \epsilon \leq 1$; the lower inequality stipulates that one lobe may not be completely buried in the other, while the upper inequality requires lobes to be in contact. We define the orientation vector ${\bm S}_i$ of peanut $i$ to be the unit vector whose origin is the center of the hydrophobic lobe and which points diametrically away from the center of the hydrophilic lobe. Neighboring peanuts interact via hard-core excluded volume interactions -- nothing may approach closer than $R_0$ (resp. $R_1$) to the center of each hydrophobic (resp. hydrophilic) lobe -- and via a short-ranged, pairwise interaction modeling the attraction between solvated nanoscale hydrophobes~\cite{hydrophobe} (we do not consider solvent explicitly in our model). We assume the hydrophobic interaction to be attenuated on a scale of approximately 5 nm, and impose an attractive interaction between the hydrophobic lobes of peanuts $i$ and $j$ of the form
\beq
\label{pot}
U_{ij}= \epsilon_{\rm b} \Theta(r_{\rm c}-r_{ij}) A(\theta_{ij}) A(\theta_{ji}) \left({\cal L}_{\alpha}(\hat{r}_{ij}) - {\cal L}_{\alpha}(\hat{r}_{\rm c})  \right).
\eeq
Here $\epsilon_{\rm b}$ is a binding energy; $\hat{x} \equiv x/R_0-1$ is a shifted and scaled distance; $r_{ij}$ is the distance between the centers of the hydrophobic lobes of $i$ and $j$, respectively C$_i$ and C$_j$; $r_{\rm c} \equiv 2.5R_0$ is a cutoff length; and ${\cal L}_{\alpha}(x) \equiv 4 (x^{-2 \alpha} - x^{-\alpha})$ is a generalized Lennard-Jones function. We take $\alpha=15$ to ensure an attraction of sufficiently short range. In simulations we varied the attractive binding energy $\epsilon_{\rm b}$ between limits of $4 \kt$ and $8 \kt$: corresponding forms of the radial component of the hydrophobic interaction potential are plotted in Fig.~\ref{figsupp2}(a). In experiment, variation of the strength of the hydrophobic driving force may be achieved by variation of temperature or solvent composition.

The factors $A$ in Equation~(\ref{pot}) parameterize an angular modulation of the attractive interaction, with $\theta_{ij}$ being the angle between ${\bm S}_i$ and the vector pointing from  C$_i$ to C$_j$. This modulation enforces the tendency of two hydrophobes of girth exceeding 1 nm to maximize their surface-to-surface contact~\cite{hydrophobe}. Its form is given and derived in the Appendix. 

We performed simulations of collections of peanuts of given geometry, defined by their values of $R_1/R_0$ and $\epsilon$, for a range of values of the hydrophobic attraction strength $\epsilon_{\rm b}$. We used 1000 peanuts in a periodically-replicated cubic simulation box of side 64$R_0$, corresponding to fixed mole fraction of colloid (this choice reflects the low concentrations that we intend to use in experiments; in exploratory simulations we verified that the phases seen here also form at several other choices of (low) concentration). The bulkiest $(R_1/R_0=2,\epsilon=1)$ and most compact $(R_1/R_0=0.2,\epsilon=0.8)$ peanuts we considered occupied volume fractions of about 14\% and 1.5\%, respectively (see Appendix). Starting from configurations in which peanuts were randomly mixed and oriented, we evolved each system according to the version of the virtual-move Monte Carlo algorithm~\cite{vmmc} described in the Appendix of Ref.~\cite{vmmc2}. This algorithm is designed to mimic an overdamped dynamics by making collective moves of particles according to the potential energy gradients, or forces, they experience; such collective moves are neglected by standard single-particle Monte Carlo algorithms. In brief, one particle is selected and subjected to a translation or a rotation. If changes in pairwise potential energies between that particle and its neighbors are favorable then the chosen particle moves independently; if not, neighbors are recruited iteratively and experience a collective displacement or translation. The acceptance rate for each move is chosen to preserve detailed balance and to reflect, in an approximate fashion, the (short-ranged) hydrodynamic drag suffered by the collective body. We scaled collective translation acceptance rates by the reciprocal of an approximate hydrodynamic radius of the moving body, appropriate for its instantaneous direction of motion~\cite{vmmc} (identical for forward and reverse moves, as required to satisfy detailed balance), and scaled acceptance rates for rotations by the cube of a similar factor, accounting for an aggregate's instantaneous axis of rotation (identical for forward and reverse rotations). While these damping factors are approximate, control of aggregate diffusion constant scalings -- and particularly enforcement of the notion that an aspherical body does not translate equally rapidly in all directions --  provides one advantage of this method over conventional integration of overdamped equations of motion. We drew translation magnitudes from a uniform distribution with maximum $0.3R_0$~\footnote{Our simulation results were qualitatively unchanged upon varying the maximum displacement from $0.05 R_0$ to $0.6 R_0$. Additionally, we also observed the phases reported here to form when we used only single-particle moves (although some dynamical pathways involving the collisions of large aggregates were suppressed).}, and drew rotation angles from a uniform distribution with maximum $13.6^{\circ}$ (rotations were performed about a randomly-chosen axis through the midpoint of the line joining the centers of hydrophobic and hydrophilic lobes). This ability to make large trial translations and rotations of individual particles in the face of attractions and repulsions that vary rapidly with distance and angle is not shared by straightforward numerical integration schemes. Based on the diffusivity in water of a body of radius $50$ nm, we estimate that our basic simulation timestep (an average of one attempted virtual move per particle) corresponds to $\sim 10^{-6}$ s. Our longest simulations exceeded $10^7$ timesteps, implying that we probe `real' timescales in excess of a second. 

\begin{figure*}[ht]
\includegraphics[width=\linewidth]{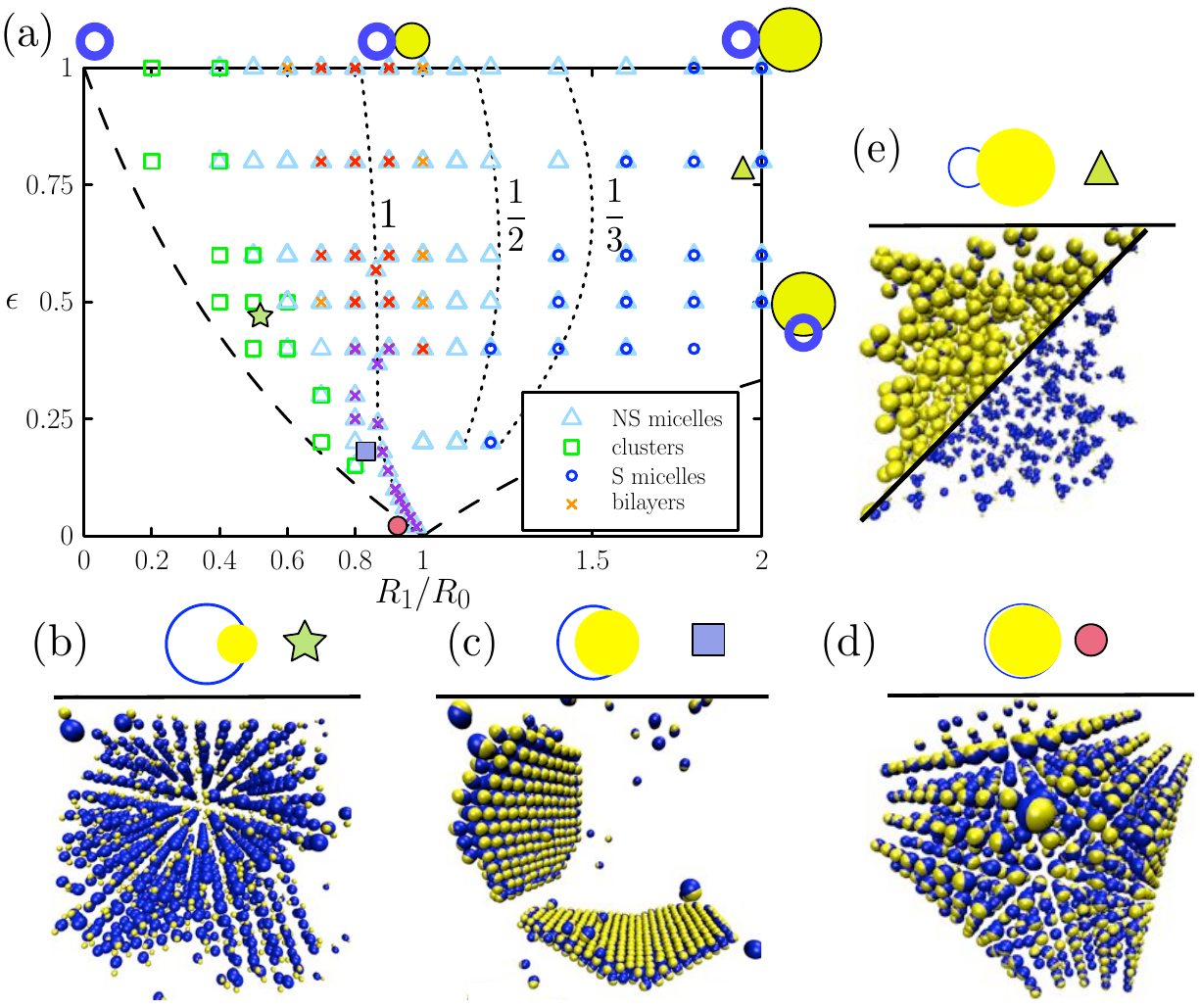} 
\caption{\label{figone} Kinetic phase diagram identifying products of self-assembly. (a) Kinetic phase diagram in the space of $\epsilon \equiv \delta/(R_0+R_1)$ and $R_1/R_0$ identifying the products found following dynamic simulations of peanuts with specified geometries (see text for classification of bilayers, micelles and clusters). Examples of such products: (b) crystalline cluster at thermodynamic state $(R_1/R_0,\epsilon,\epsilon_{\rm b}) = (0.5,0.5,4 \kt)$ (peanuts shown reduced in size); (c) bilayers at $(R_1/R_0,\epsilon,\epsilon_{\rm b}) = (0.9,0.2,5 \kt)$; (d) faceted polyhedron at state $(R_1/R_0,\epsilon,\epsilon_{\rm b}) = (0.96,0.04,4.75\kt)$ (peanuts shown reduced in size); and (e) spherical micelles at $(R_1/R_0,\epsilon,\epsilon_{\rm b}) = (2,0.8,8 \kt)$ (bottom right shows hydrophilic lobes reduced in size). Contours of constant packing parameter (values indicated) are dotted lines; the dashed lines indicate the two branches satisfying $\epsilon = |R_1 -R_0|/(R_1+R_0)$. Below these lines one lobe of the `peanut' is completely buried within the other.}
\end{figure*}

Our results are summarized by the `kinetic phase diagram' of Fig.~\ref{figone}(a). This diagram identifies those self-assembled products, whether equilibrated or kinetically trapped, accessible to dynamical simulation starting from well-mixed initial conditions. We expect by extension that such products will be accessible to experiments starting from similar conditions. The horizonal and vertical axes of this diagram label the quantities $R_1/R_0$ and $\epsilon$, respectively; peanuts with small (resp. large) hydrophilic lobes are found to the left (resp. right) of the diagram. We identify regions of compact clusters (squares; see Fig.~\ref{figone}(b)), nonspherical micelles (triangles), spherical micelles (circles; see Fig.~\ref{figone}(e)) and bilayers (crosses; see Fig.~\ref{figone}(c)). Our classification scheme and a summary of the computational resources we deployed are discussed in the Appendix. We find these phases to be localized in peanut shape space largely in accord with a simple estimate of peanuts' molecular packing parameters~\cite{israelachvili1976tsa,nagarajan2002mpp} (see Appendix): we have labeled the diagram with lines of packing parameter equal to 1 (the regime in which one expects bilayers); $1/2$ (the cylindrical micelle regime); and $1/3$ (the spherical micelle regime). 

\begin{figure*}[ht]
\includegraphics[width=\linewidth]{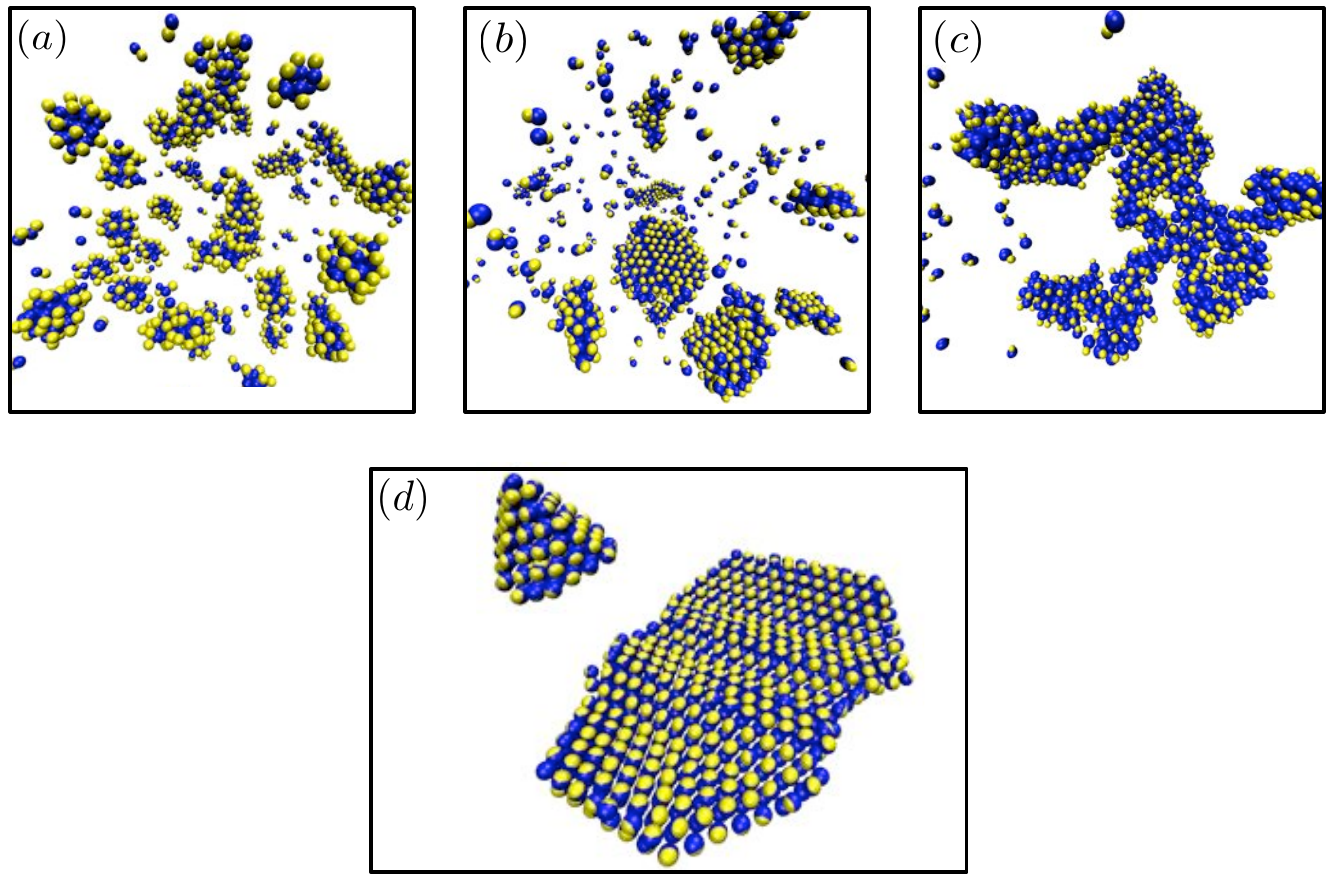} 
\caption{\label{figsupp} Additional configurations. (a) Micelles of various morphologies found at thermodynamic state $(R_1/R_0,\epsilon,\epsilon_{\rm b}) = (1,0.8,5.5 \kt)$; (b) coexisting bilayers and micelles at state $(R_1/R_0,\epsilon,\epsilon_{\rm b}) = (0.8,0.5, 5 \kt)$; (c) disordered wormlike micelle at state $(R_1/R_0,\epsilon,\epsilon_{\rm b}) = (0.6,0.8,5 \kt)$; and (d) coexisting polygon and bilayer at state $(R_1/R_0,\epsilon,\epsilon_{\rm b}) = (0.96,0.04,4.75 \kt)$. 
}
\end{figure*}

The phase most interesting from the materials scientist's perspective is perhaps the bilayer, because extended two-dimensional structures that do not require templating by a substrate are attractive candidates for device fabrication. Our observation that amphiphilic nanoscale peanuts can form bilayers, and that bilayer formation is localized to a specific region of peanut shape space, will facilitate the experimental search for this phase. We also observed bilayers to coexist with faceted polyhedra composed of peanuts arranged with a high degree of local order (see Fig.~\ref{figone}(d) and Fig.~\ref{figsupp}(d)). Like micelles, and unlike extended clusters, polyhedra present hydrophilic surfaces to solution and are consequently size-limited. We do not know why bilayers and polyhedra coexist: they may represent comparable minima of free energy, or one may embody a particularly stable kinetic trap. This coexistence  -- which resembles simultaneous micellization and phase separation~\cite{sciortino2009phase} -- will be the focus of future work, but its observation nonetheless suggests that experimental realizations of this system might achieve multipotent self-assembly, wherein components form coexisting, ordered phases of strikingly different symmetry. Some protein complexes appear to achieve such multipotent assembly by executing conformational changes~\cite{whitelam2008icf}. 

\begin{figure}[ht]
\includegraphics[width=0.7\linewidth]{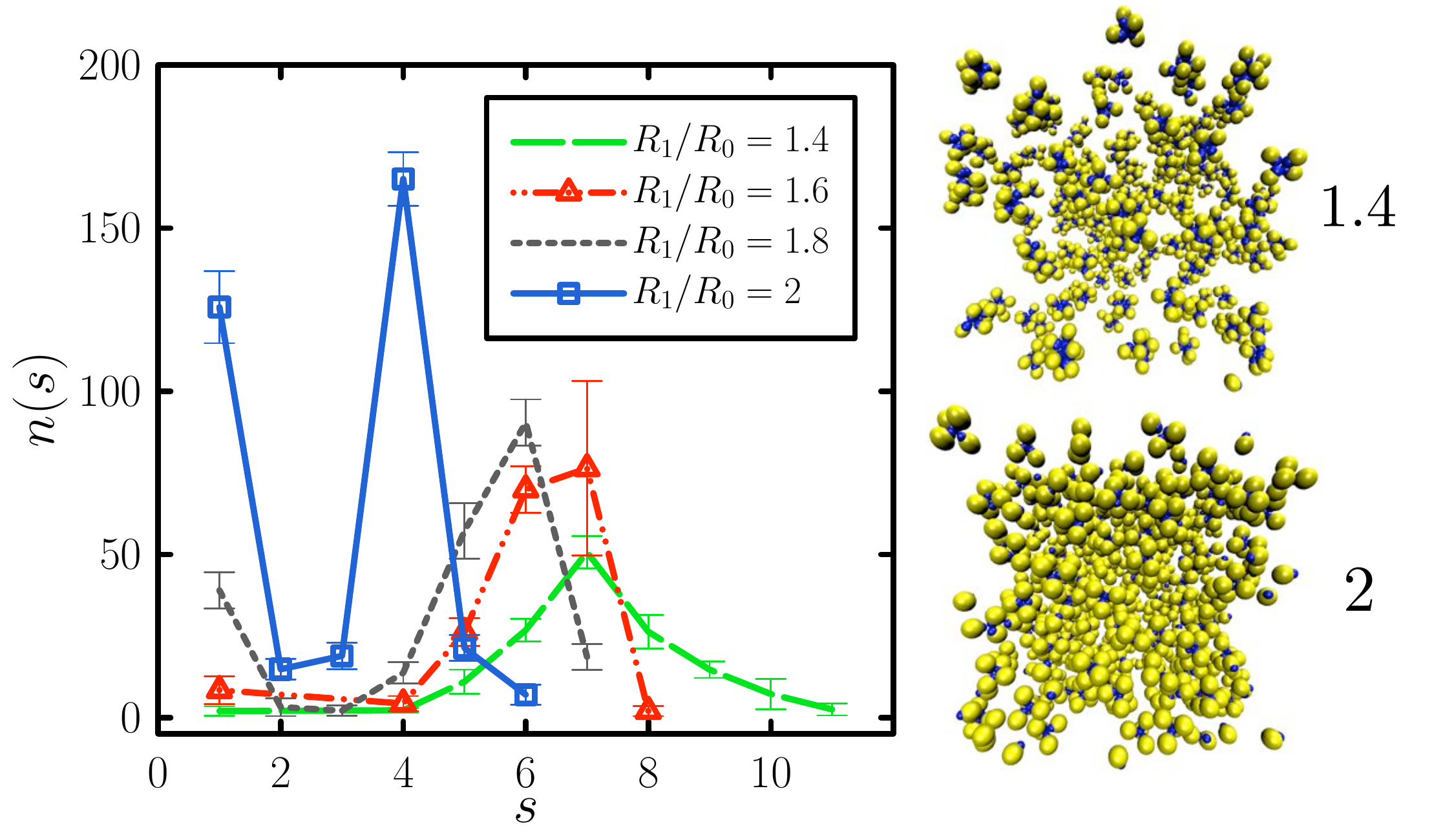} 
\caption{\label{figtwo} Micelle polydispersity. Number of aggregates of size $s$, $n(s)$, for micelle-forming peanuts satisfying $\epsilon=1/2$ and $\epsilon_{\rm b} = 8 \kt$. Varying $R_1/R_0$, which is achievable synthetically, controls micelle polydispersity. Each line was obtained using 10 independent simulations evolved for $\sim 5 \times 10^6$ timesteps.}
\end{figure}

Of the other phases observed, micelles are found abundantly for many peanut shapes, especially when the thermodynamic driving force for association is large and the formation of ordered structures like bilayers is hindered by kinetic traps. Micelle polydispersity may be controlled by varying $R_1/R_0$ (Fig.~\ref{figtwo}), suggesting a route to the synthesis of size-controlled nanometer scale assemblies. In addition, cluster formation is possible when the hydrophilic lobe is very small. Peanuts whose shapes lie on the left branch of the line $\epsilon = |R_1 -R_0|/(R_1+R_0)$ in Fig.~\ref{figone}(a) are isotropically attractive spheres, and form close-packed, crystalline clusters. Peanuts with small hydrophilic lobes also form aggregates whose innermost particles make 12 pairwise contacts, and extended crystalline order is possible even when the hydrophilic lobe is moderately sized (see Fig.~\ref{figone}(b)).

\begin{figure*}[ht]
\includegraphics[width=\linewidth]{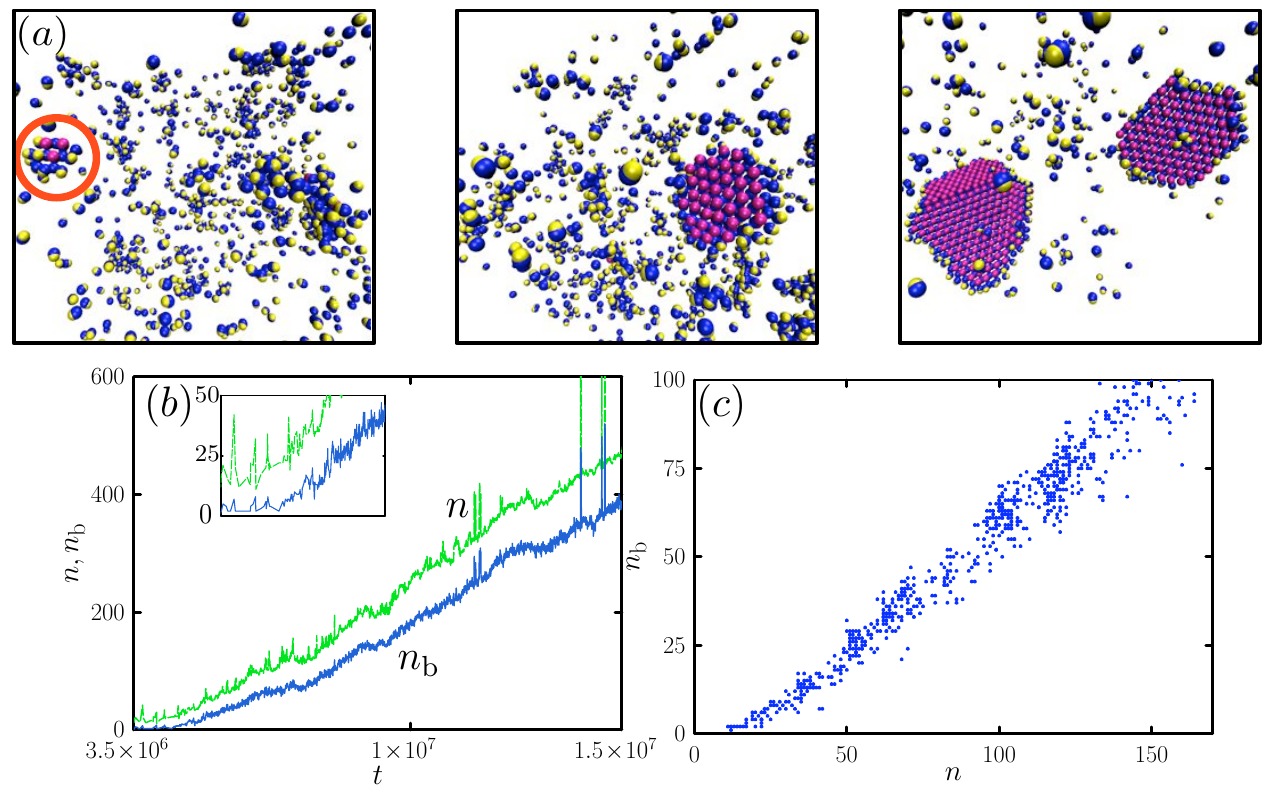} 
\caption{\label{figthree} Bilayer self-assembly from one simulation at thermodynamic state $(R_1/R_0,\epsilon,\epsilon_{\rm b}) = (0.9,0.2,5 \kt)$. (a) Time-ordered configuration snapshots. Left ($t\approx 2.1 \times 10^6$): a micelle establishes a bilayer-like configuration, but fails to grow and eventually evaporates. Bilayer-like particles (see Appendix) are shown in pink. Center ($t\approx 6.8 \times 10^6$): a similar structure has undergone a collective rearrangement and attained a critical size, and as a result grows readily. Right ($t\approx 1.6 \times 10^7$): it is eventually joined by a second bilayer; these grow until they possess about 800 of the 1000 peanuts in the simulation box. (b) As a function of number of simulation timesteps $t$ we plot number of particles $n$ and number of bilayer-like particles $n_{\rm b}$ comprising the structure seen in the center panel (which occasionally merges with the second bilayer seen in the right panel). Inset: early-time behavior. These quantities are plotted in (c) with $t$ as a parameter.}
\end{figure*}

\begin{figure*}[ht]
\includegraphics[width=\linewidth]{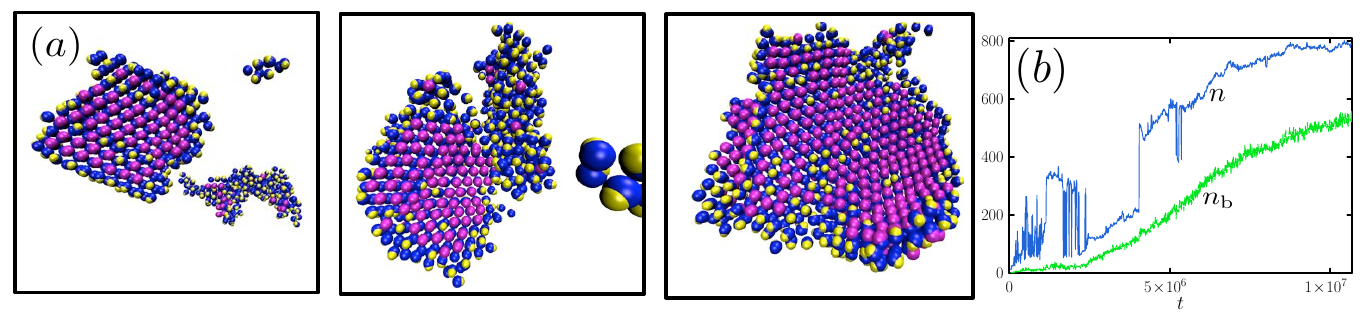} 
\caption{\label{figfour} Bilayer self-assembly from one simulation at thermodynamic state $(R_1/R_0,\epsilon,\epsilon_{\rm b}) = (0.8,0.25,4.75 \kt)$. (a) Configurations at $t=(4,4.5,9) \times 10^6$ (left to right): a bilayer and a micelle merge, with the former eventually consuming the latter. Bilayer-like particles are shown in pink. Only aggregates possessing more than 5 constituents are shown. (b) Size $n$ and number of bilayer-like particles $n_{\rm b}$ comprising the most bilayer-like aggregate in the simulation box, as a function of $t$.}
\end{figure*}

While considerations of packing predict the phases we have found, with the exception of faceted polyhedra, they give little insight into the complex dynamics of assembly we have observed. We focus on the hierarchical dynamics associated with peanut state $(R_1/R_0,\epsilon,\epsilon_{\rm b})$ = $(0.9,0.2,5 k_{\rm B}T)$. First, fluctional micelles appear. These can rearrange in a collective fashion to form a flattish, metastable proto-bilayer composed of about 7 close-packed peanuts per face, plus attendant edge particles (see Fig.~\ref{figthree}(a), left, circled). Such nuclei do not grow immediately, but do so only after an additional collective rearrangement that sees an abrupt increase in the number of bilayer-like particles comprising their core (see Fig.~\ref{figthree}(b,c)). As the nucleus exceeds a critical size, bilayer growth proceeds readily; the largest bilayers we observed in systems of 1000 peanuts exceeded 800 particles in size (in exploratory simulations of 6000 peanuts we observed bilayers comprising in excess of 2000 peanuts). The time of the appearance of the first bilayer-like nucleus exceeding 50 particles in size is long and drawn from a broad distribution (from 20 independent simulations of $\sim 2.5 \times 10^7$ timesteps we observed the appearance of such structures in 17 cases; of this set, mean nucleation time was $1.2 \times 10^7$ timesteps, with a standard deviation of $9\times10^6$ timesteps). We have not explored in detail the dynamics of bilayer formation at other points in peanut shape space, but we present one example in Fig.~\ref{figfour}. Here a bilayer and a micelle merge before the former grows at the expense of the latter.

Finally, we note that cluster formation by `peanuts' whose hydrophilic lobes are of vanishing size proceeds via `two-step' crystallization~\cite{wolde1997epc,chen2008avb,vanmeel2008tsv,fortini2008cag}: first, dense amorphous blobs nucleate from vapor; second, crystalline order nucleates within these blobs (not shown). We observed a similar dynamics (not shown) for crystal-forming peanuts (see Fig.~\ref{figone}(b)) possessing hydrophilic lobes of moderate size. It would make for an interesting theoretical study to determine the dynamics of assembly as one proceeds into the regime of increasing hydrophilic lobe size: what is the nature of cluster formation as the crystal structure becomes suppressed or thermodynamically disfavored with respect to the dense amorphous phase? Clusters in our simulations generally coalesce upon contact, suggesting that in experiment such assemblies would not be soluble, and would not be of great practical interest.

We have demonstrated that dynamical simulation of a simple model of nanoscale amphiphilic peanut-shaped colloids generates phases of clusters, micelles and bilayers in accord with expectations based on simple ideas of packing. We have also found certain ordered structures and observed complex dynamical mechanisms that cannot be so explained. We expect this model and its behavior to be prototypical of a class of amphiphilic structures now being synthesized in large quantities, and therefore to help guide the experimental search for new, functional nanostructured materials. Future theoretical work involving this model will focus on further quantifying the dynamical pathways observed in this study; determining the effect upon assembly of liquid-liquid interfaces~\cite{cheung2009ini}; and studying the design potential of multi-lobed peanut generalizations.\\

{\em Note:} In a recent study, Miller and Cacciuto explored the self-assembly of spherical amphiphilic particles using molecular dynamics simulation~\cite{miller2009hsa}. Interestingly, despite the differences in interaction range and particle shape between their model and ours, those authors also observed the coexistence of bilayers and faceted polyhedra. This implies that multipotent assembly of this nature is not dependent upon fine details of particle-particle interactions, but can be understood on more general grounds. Furthermore, it may be that such multipotent assembly has already been seen in experiment: one of the referees of this paper pointed out that surfactant bilayers can form icosahedral structures in salt-free cationic solution~\cite{dubois2001self}. One possible origin of these structures is the electrostatic force~\cite{vernizzi2007faceting}; it is also conceivable, based on our results and the results of Ref.~\cite{miller2009hsa}, that their existence might be explained in purely geometrical terms.

\section{Acknowledgements}
This work was performed at the Molecular Foundry, Lawrence Berkeley National Laboratory, and was supported by the Director, Office of Science, Office of Basic Energy Sciences, of the U.S. Department of Energy under Contract No. DE-AC02 -- 05CH11231. We thank Sander Pronk for a critical reading of the manuscript, and Andrea Pasqua, Lutz Maibaum and Phillip Geissler for useful discussions.

\break

\section{Appendix}

\begin{figure*}[ht]
\includegraphics[width=\linewidth]{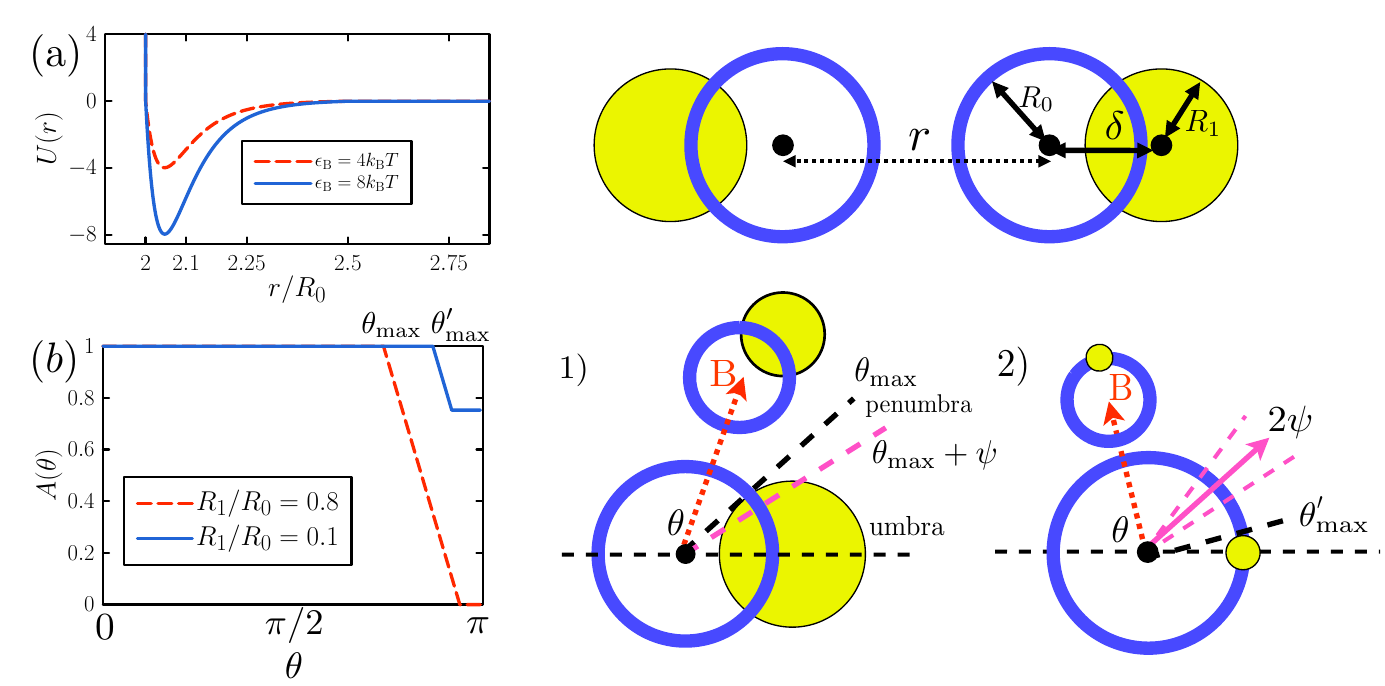} 
\caption{\label{figsupp2} Peanut-peanut interaction potential. (a) Peanut geometry with hydrophobic (resp. hydrophilic) lobe colored blue (resp. yellow), together with the radial component of the blue-blue interaction potential. (b) Angular modulation function, $A(\theta)$, designed to account for the degree of contact made by hydrophobic lobes of adjacent peanuts (see Equation~\ref{ang_pot}). $\theta$ is the plane angle between the peanut orientation vector and the vector pointing from the center of the hydrophobic lobe (blue) to the center of a similar body, B (shown reduced in size). We plot $A(\theta)$ for peanuts of geometry 1) $(R_1/R_0,\epsilon) = (0.8,0.85)$ and 2) $(R_1/R_0,\epsilon) = (0.1,0.85)$. `Penumbra' and `umbra' identify the regions for which, respectively, partial and total occlusion of the blue lobe is caused by the yellow lobe. See text for details.}
\end{figure*}

\noindent
{\em Angular modulation of the hydrophobic interaction}. The factors $A$ in Equation~(\ref{pot}) have the following origin (refer to Fig.~\ref{figsupp2} for geometry). We assume the strength of the hydrophobic attraction on the lengthscales we consider to be proportional to the contacting surface area between two hydrophobic bodies~\cite{hydrophobe}. We define two surfaces to be `in contact' if they lie no further than 5 nm apart, reflecting our assessment of the likely range of the hydrophobic force. If two spheres of radius $R_0 \approx 50 $ nm are placed in contact, the plane half-angle subtended by the limit of the `contacting' surface area at the center of either sphere is $\psi \equiv \arccos(1-5/(2 \cdot 50)) \approx 18.1^{\circ}$ (peanuts are azimuthally symmetric about their orientation vector, and so we consider only plane angles, as sketched in Fig.~\ref{figsupp2}). Hence we argue that the hydrophobic lobe of peanut $i$ presents its maximum possible surface area to a similar body B if the angle between ${\bm S}_i$ and the vector pointing from C$_i$ to the center of B is {\em not} within an angle $\psi$ of the line joining C$_i$ to the nearest surface obstruction. 

We next define the angle $\theta_{\rm max}$ to be the angle between ${\bm S}_i$ and the line joining C$_{i}$ to the largest obstruction provided by each hydrophobic lobe's hydrophilic partner. This angle is given by geometrical considerations. The angle between ${\bm S}_i$ and the line joining C$_{i}$ to the intersection of hydrophobic and hydrophilic lobes is $\theta_{\rm intersec} = \pi - \arccos(\ell/R_0)$, where $\ell \equiv \delta/2-(R_1^2-R_0^2)/(2 \delta)$ (note that $\ell$ may be negative). The angle between ${\bm S}_i$ and the line joining C$_{i}$ to the greatest diameter presented by its hydrophilic partner is $\theta_{\rm girth} = \pi - \arctan(R_1/\delta)$. We take $\theta_{\rm max} = \min (\theta_{\rm intersec},\theta_{\rm girth})$, {\em unless} the largest diameter of the hydrophilic lobe is buried in its hydrophobic counterpart (corresponding to $R_1<R_0$ and $\ell>\delta$); in this case we take $\theta_{\rm max} = \theta_{\rm intersec}$.

To construct the function $A(\theta)$ we argue as follows. Let $\theta$ be the angle between ${\bm S}_i$ and the line joining C$_{i}$ to the center of a similar body B. For $\theta=0$ the peanut $i$ presents its maximum possible hydrophobic surface area to B. Conversely, the greatest obstruction of the hydrophobic lobe caused by the hydrophilic lobe occurs when $\theta=\pi$, i.e. when the body B approaches the peanut from its hydrophilic side. This obstruction can be total, if the hydrophilic lobe is sufficiently large, or less than total, if the hydrophilic lobe is small and/or well-buried within its hydrophobic partner. We take the smallest possible `signal' presented by the hydrophobic surface of peanut $i$ to the body B to be $A_{\rm min}= \max(0,1-(\pi - \theta_{\rm max})/\psi)$. Finally, we assume for simplicity that the signal interpolates linearly with $\theta$ between its largest and smallest values, i.e $A(\theta) \propto -\theta$ if B lies in the penumbra cast by the hydrophilic lobe (when $\theta $ exceeds $ \theta_{\rm max} -\psi$). These arguments imply the following piecewise linear form for the angular modulation function in Equation~(\ref{pot}):
\beq
\label{ang_pot}
A(\theta) = \left\{
 \begin{array}{ll} 
 1 &( \theta <\phi)  \\
 \max(A_{\rm min},1-\frac{1}{2 \psi}(\theta-\phi) ) & (\theta \geq\phi),
 \end{array}
 \right.
\eeq
where $\phi \equiv \theta_{\rm max}- \psi$. In Fig.~\ref{figsupp2}(b) we plot $A(\theta)$ for two peanut geometries. We note that for the nanoscale particles we have studied here, identification of the length and angular scales over which the hydrophobic effect operates reveals that multibody forces between particles are neither required nor warranted.

We note that our estimate of the range of the hydrophobic attraction is rough (see e.g.~\cite{lin2005measurement}). If our assessment of the range of the hydrophobic force is inaccurate, then our results will describe nanoparticles whose size is not the 50 nm assumed here. We have not performed systematic simulations using different ranges of attraction, but exploratory simulations of peanuts whose hydrophobic lobes' radial attractions have a square well form of range equal to the lobe radius can form at least one phase not accessible to particles possessing the shorter range of attraction studied in the text. This phase consists of micelle-like blobs (of many particle diameters in girth) possessing dense, liquidlike interiors. These blobs can grow through their mutual coalescence and subsequent restructuring. We expect therefore that small nanoparticles (of girth, say, $\sim 5$ nm) can form phases that larger nanoparticles (of size, say, $\sim 50$ nm) cannot. \\

\noindent
{\em Peanut volume.} The volume presented to solvent by two fused spheres of radii $R_0$ and $R_1$ whose centers are $\delta$ apart $(|R_1 -R_0| < \delta \leq (R_1+R_0))$ is given from geometrical considerations by
\bea
V_{\rm peanut} &=& \frac{4}{3} \pi (R_0^3+R_1^3) 
- \pi \int_{\ell}^{R_0} dx (R_0^2-x^2) \nonumber \\
&-&\pi \int_{\delta-\ell}^{R_1} dx (R_1^2-x^2) \nonumber \\
&=&\frac{(1 + \epsilon)^2}{12 \epsilon} \pi \Sigma \left(3 \Delta^2 + 
    \epsilon (2-\epsilon) \Sigma^2  \right).
\eea
Here $\epsilon \equiv \delta/(R_0+R_1)$, $\ell \equiv \delta/2-(R_1^2-R_0^2)/(2 \delta)$, $\Sigma \equiv R_0+R_1$ and $\Delta \equiv R_0-R_1$. This formula may be used to calculate the fraction of the simulation box occupied by peanuts of arbitrary geometry.\\

\noindent
{\em Computational details}. To obtain Fig.~\ref{figone}(a) we performed simulations for times sufficient to observe what we believe to be steady-state behavior at each location in peanut shape space. In the regime $R_1/R_0\geq 1.2$ (the micelle-forming regime) this is straightforward: for each point in the $(R_1/R_0,\epsilon)$ plane we performed one simulation at each of $\epsilon_{\rm b}=(5,6,7,8) k_{\rm B}T$ for $\sim 5 \times 10^6$ timesteps (60 CPU hours per simulation). In the cluster-forming regime of $R_1/R_0\leq 0.4$  we performed 5 independent simulations at each labeled point for each value of  $\epsilon_{\rm b}=(4,4.5,5,5.5) k_{\rm B}T$, also for $\sim 5 \times 10^6$ timesteps each (60 CPU hours per simulation). In the intermediate regime $0.4< R_1/R_0< 1.2$ we found some bilayers to form only after very long times. We therefore performed 5 independent simulations  of $\sim 5 \times 10^6$ timesteps each (60 CPU hours per simulation) at each point for each value of $\epsilon_{\rm b}=(4,4.5,4.75,5,5.5) k_{\rm B}T$, and in addition performed three further independent simulations of $\sim 2.5 \times 10^7$ timesteps (300 CPU hours per simulation) at each point for each of the five values of $\epsilon_{\rm b}$. Along the line $R_1/R_0=1$ we performed three additional simulations of  $\sim 2.5 \times 10^7$ timesteps for each value of $\epsilon_{\rm b}=(6,6.5) k_{\rm B}T$. We can of course not rule out the likelihood of nucleation and growth of structures on timescales longer than those we probed, nor rule out the appearance of other interesting ordered structures at values of $\epsilon_{\rm b}$ that we did not consider. To obtain Fig.~\ref{figtwo} we performed 10 additional simulations of  $\sim 5 \times 10^6$ timesteps at each of the four thermodynamic states shown (60 CPU hours per simulation). To obtain nucleation statistics for bilayers similar to those seen in Fig.~\ref{figthree}) we performed 20 additional independent simulations of $\sim 2.5 \times 10^7$ timesteps (300 CPU hours per simulation) at thermodynamic state $(R_1/R_0,\epsilon,\epsilon_{\rm b}) = (0.9,0.2,5 \kt)$.\\

\noindent
{\em Classification of self-assembled structures}. We classified the products of peanut self-assembly by analyzing the final configuration of each simulation as follows. We define contacting peanuts to be those possessing a pairwise energy of interaction of $-\kt$ or less. We define an `aggregate' to be a contiguous set of contacting particles. We identified locations in peanut shape space to contain `clusters' (compact structures with some degree of close packing) if any particle possessed 12 contacts. We identified micelles by calculating for each aggregate the number ${\cal M} \equiv  N_{\rm a}^{-1} \sum_{i=1}^{N_{\rm a}} {\bm S}_i \cdot ({\bm r}_{\rm CM} - {\bm r}_i)$. Here $N_{\rm a}$ is the number of peanuts comprising the aggregate, ${\bm r}_{\rm CM}$ is the center of mass of the aggregate, and ${\bm r}_i$ is the position of aggregate constituent $i$ (all coordinates are corrected for periodic boundaries). If any location in shape space contained at least three aggregates of size three or greater that scored ${\cal M} \geq0.9$ then we considered that state to contain spherical micelles, and marked that state with a circle. Any location with three or more aggregates of size three scoring $0.5 \leq {\cal M} < 0.9$ is considered to contain nonspherical micelles, and was marked with a triangle. Bilayer order was identified on the basis of the order parameter ${\cal B} \equiv  2 \left(N_{\rm a} ( N_{\rm a}-1)\right)^{-1} \sum_{(ij)} ({\bm S}_i  \cdot {\bm S}_j) ^2$, where the sum runs over all particle pairs in the aggregate (${\cal B}=\frac{1}{3}$ for collections of randomly-oriented peanuts). Configurations possessing an aggregate of (25,50,200) constituents or more having a score of  ${\cal B} \geq 0.4$ were marked with an (orange, red, purple) cross. In Figs.~\ref{figthree} and~\ref{figfour} we show for certain aggregates the number of their constituents possessing local bilayer-like order, $n_{\rm b}$. To compute this quantity we formed the dot product of the orientation vector of a given particle and each of its neighbors' orientation vectors. We define a `bilayer-like' particle as one possessing 3 or more neighbors with which it has orientation-orientation dot product $>0.8$, and 1 or more neighbors with which it has dot product $<-0.8$. This order parameter is not useful for examining clusters, whose internal particles associate in an orientationally disordered way and can test `false positive' for local bilayer ordering. However, it does successfully distinguish between regions of local order and disorder in the bilayer-forming regime of peanut shape space. For most conditions considered it is possible to find amorphous structures answering to none of the above descriptions. This is most often true when the hydrophobic driving force is very strong, and kinetically frustrated aggregates form. We have ignored these structures, focusing on the ordered products we have described. While the order parameters and the threshold numbers we have applied are arbitrary, and the extent of each regime varies as the structural classification criteria are varied, we consider the trends we have identified to be representative of the nature of the self assembly we observed. Some additional structures are shown in Fig.~\ref{figsupp}. We used VMD~\cite{vmd} to render simulation configurations in this paper.\\

\noindent
{\em Packing parameter.} Considerations of packing have been used with success to predict the phases formed by molecular surfactants~\cite{nagarajan2002mpp, israelachvili1976tsa}. We can define a packing parameter for an aggregate as $P_{\rm agg}\equiv V_{\rm agg}/(a_{\rm agg} L_{\rm agg})$, where $V_{\rm agg}$ is aggregate volume, $a_{\rm agg}$ its surface area, and $L_{\rm agg}$ a characteristic length. For a cylinder of radius $R$ and length $\ell$, for example, $V_{\rm cylinder}=\pi R^2 \ell$, $L_{\rm cylinder}=R$, and $a_{\rm cylinder}=2 \pi R \ell$, giving $P_{\rm cylinder}=1/2$. Analogously, $P_{\rm sphere}=1/3$ and $P_{\rm bilayer} =1$. If we now assume such shapes to be composed of amphiphilic components, then, loosely, $a_{\rm agg}$ corresponds to the total area presented to solvent by hydrophilic groups, while $V_{\rm agg}$ and $L_{\rm agg}$ are the volumes and characteristic lengths of the hydrophobic units. If one assumes that $P_{\rm agg}$ can be estimated from the geometry of a {\em single} amphiphile (see, however, Ref.~\cite{nagarajan2002mpp}), then values of $P_{\rm single} \equiv P$ near $1$, $1/2$ or $1/3$ suggest self-assembly of bilayers, cylindrical micelles or spherical micelles, respectively. We can make a rough estimate of $P$ for an individual peanut. If we assume that hydrophobic and hydrophilic lobes touch but do not interpenetrate, then $V=\frac{4}{3} \pi R_0^3$, $L=2R_0$ and $a=\pi R_1^2$. We thus have $P = \frac{2}{3} (R_0/R_1)^2$. To this approximation, therefore, contours of constant $P$ are vertical lines on Fig.~\ref{figone}(b). Furthermore, $P=1 \implies R_1/R_0 = \sqrt{2/3} \approx 0.82$ and $P=1/3 \implies R_1/R_0 = \sqrt{2} \approx 1.41$. From Fig.~\ref{figone}(b) we indeed find bilayers and spherical micelles in the corresponding regimes. A more refined estimate based on peanut geometry alone may be made by accounting for the volume of the hydrophobic lobe buried within the hydrophilic lobe. We have 
\beq
V= \frac{4}{3} \pi R_0^3- \pi \int_{\ell}^{R_0} dx (R_0^2-x^2),
\eeq
$L=R_0+\ell$, and $a= \pi R_1^2$ (valid when the hydrophilic lobe's greatest diameter is not buried within the hydrophobic lobe; if this is {\em not} true, then $a= \pi (R_0^2-\ell^2)$). Contours of $P=1$, $P=1/2$ and $P=1/3$ to this approximation are plotted in Fig.~\ref{figone}(a).

\break

\section{References}
\bibliography{bib}
\break

\section{Figures}

\break

\break

\break

\break

\end{document}